\documentclass[pre,aps,superscriptaddress,amsmath,showpacs,floatfix]{revtex4}
\usepackage{graphicx}
\usepackage{epsfig}
\usepackage{amssymb}
\begin{document}
\title{Membrane fluctuations near a plane rigid surface}
\author{Oded Farago}
\affiliation{Department of Biomedical Engineering, Ben Gurion University,
  Be'er Sheva 84105, Israel}
\begin{abstract}

We use analytical calculations and Monte Carlo simulations to
determine the thermal fluctuation spectrum of a membrane patch of a
few tens of nanometer in size, whose corners are located at a fixed
distance $d$ above a plane rigid surface. Our analysis shows that the
surface influence on the bilayer fluctuations can be effectively
described in terms of a uniform confining potential that grows
quadratically with the height of the membrane $h$ relative to the
surface: $V=(1/2)\gamma h^2$. The strength $\gamma$ of the harmonic
confining potential vanishes when the corners of the membrane patch
are placed directly on the surface ($d=0$), and achieves its maximum
value when $d$ is of the order of a few nanometers. However, even at
maximum strength the confinement effect is quite small and has
noticeable impact only on the amplitude of the largest bending mode.
\end{abstract}
\maketitle

\section{Introduction}
\label{sec:intro}

Fatty acids and other lipids are essential to every living
organism. Because of their amphiphilic nature, they spontaneously
self-assemble into bilayer membranes that define the limits of cells
and serve as permeability barrier to prevent proteins, ions and
metabolites from leaking out of the cell and unwanted toxins leaking
in \cite{Lipowsky_Sackmann}. In euokaryotic cells, membranes also
surround the organelles allowing for organization of biological
processes through compartmentalization. In addition, biological
membranes host numerous proteins that are crucial for the mechanical
stability of the cell, and which carry out a variety of functions such
as energy and signal transduction, communication, and cellular
homeostasis \cite{Alberts}.

An important aspect of biological membranes is that they are typically
not free but rather confined by other surrounding membranes, adhere to
other membranes, and attach to elastic networks like the cytoskeleton
and the extracellular matrix. Several model systems with reduced
compositional complexity have been designed to mimic biological
membranes. These biomimetic systems include phospholipid bilayers
deposited onto solid substrates (solid-supported
membranes)~\cite{Saldit:2005}, or on ultra-thin polymer supports
(polymer-supported membranes)~\cite{Tanaka_Sackmann:2005}. With the
aid of biochemical tools and generic engineering, supported membranes
can be functionalized with various membrane-associated
proteins~\cite{Tanaka_Sackmann:2006}. Synthetic supported membranes
with reconstituted proteins are increasingly used as controlled
idealized models for studying key properties of cellular
membranes~\cite{Girard:2007}. They provide a natural environment for
the immobilization of proteins under nondenaturating conditions and in
well-defined orientation~\cite{Salafsky:1996a}. Another attractive
application of supported membranes is the design of phantom cells
exhibiting well defined adhesive properties and receptor
densities~\cite{Salafsky:1996b}. Finally, biofunctional membranes
supported by solid interfaces (semiconductors, metals, plastics)
provide new classes of biosensors, diagnostic tools, and other
biocompatible materials~\cite{Tanaka_Sackmann:2006,Daniel:2006}.

Theoretically, the thermal shape undulations of supported membranes
have been addressed for various model systems. These model systems
include: (i) membranes that adhere to surfaces under the action of a
{\em continuum}\/ local potential \cite{Swain:2001,Mecke:2003}, (ii)
membranes pinned or tethered {\em discretely}\/ to a surface
\cite{Lin:2004,Lin:2006,Merath:2006,Merath:2007}, and (iii) membranes
supported by elastic networks of springs
\cite{Gov:2003,Fournier:2004,Fournier:2006,Thorsten:2007,Zhang:2008}. The
investigation of the latter case is largely motivated by recent
studies of simple cells (e.g., red blood cells and the lateral cortex
of auditory outer hair cells), where the supporting cytoskeleton has a
fairly well defined connectivity \cite{Gov_Safran:2004}. One feature,
missing in many previous theoretical studies, is the influence of
steric (``excluded volume'' - EV) interactions between the membrane
and its support on the elasticity and shape fluctuations of the
membrane. When these interactions are considered (see, e.g.,
refs.~\cite{Mecke:2003,Bruinsma:1994}), it is usually assumed that the
disjoining potential due to the collisions between the membrane and
the underlying surface decreases inverse-quadratically with the
distance between them. This result has been originally obtained in a
heuristic manner by Helfrich \cite{Helfrich:1978}, and later was
formulated more systematically by using a renormalization-group
approach and computer simulations (see \cite{Lipowsky} and
refs.~therein). An instructive way to understand this result is as
follows: Consider a membrane that spans a planar square frame of area
$L^2$. The Helfrich energy (to quadratic order) for a nearly-flat
membrane in the Monge gauge is given by
\begin{equation}
{\cal H} = \int\, 
\frac{\kappa}{2}\left(\nabla^2 h \right)^2d^2\vec{r},
\label{eq:helfrich} 
\end{equation}
where $\kappa$ the bending rigidity, and $h$ the height of the
membrane above the frame reference plane. Dividing the frame area into
$\Omega_0=(L/l)^2$ grid cells of microscopic area $l^2$ (where $l$ is
of the order of the thickness of the bilayer), and introducing the
Fourier transform of $h\left(\vec{r}\right)$
\begin{equation}
h_q=\frac{1}{L^2}
\int d^2\vec{r}\, h\left(\vec{r}\right)e^{i\vec{q}\cdot\vec{r}}
\ \ ;\ \ \vec{q}=\frac{2\pi}{L}\left(n_1,n_2\right),
\ n_1,n_2=-\frac{L}{2l},\ \ldots\ ,\frac{L}{2l},
\label{eq:hFourier}
\end{equation}
the Helfrich energy takes the form
\begin{equation}
{\cal H} = \frac{l^4}{L^2}\sum_{\vec{q}}\frac{\kappa}{2} q^4|h_q|^2,
\label{eq:qhelfrich}  
\end{equation}
from which (by invoking the equipartition theorem) one finds that
\begin{equation}
\langle|h_q|^2\rangle =\frac{k_BT L^2}{\kappa l^4 q^4},
\label{eq:equipartition}
\end{equation}
where $k_B$ is Boltzmann factor, $T$ the temperature, and
$q=|\vec{q}|=(2\pi/L)(n_1^2+n_2^2)^{1/2}\equiv(2\pi/L)n$. Using the
last result, one readily finds that the mean-square fluctuation of the
height increases quadratically with $L$
\begin{equation}
\Delta_0^2=\langle h\left(\vec{r}\right)^2\rangle=
\left(\frac{l}{L}\right)^4
\sum_{\vec{q}} \langle |h_q|^2 \rangle 
= \frac{k_BT}{\kappa}\frac{L^2}{(2\pi)^4}
\sum_{\vec{q}}\frac{1}{n^4}\simeq \frac{6.03k_BT}{(2\pi)^4\kappa}L^2.
\label{eq:meanhfluct}
\end{equation}

\begin{figure}
\vspace{1.5cm}
  {\centering
  \hspace{1.5cm}\epsfig{file=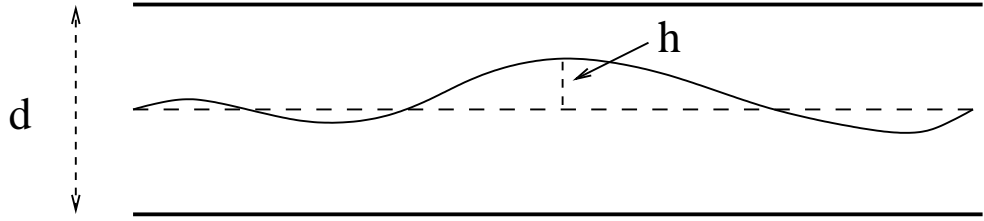,width=13cm}}
\caption{A fluctuating membrane confined between two walls which are 
separated from each other by a distance $d$. The height of the
fluctuating membrane, $h$, is measured from the mid plane between the
walls ($-d/2\leq h\leq +d/2$).}
\label{fig:twowalls_schm}
\end{figure}

Now, consider a membrane placed between two parallel walls positioned
a distance $d$ from each other, as shown schematically in
Fig.~\ref{fig:twowalls_schm}. The presence of the walls significantly
suppresses the long wavelength thermal fluctuations of the confined
membrane. At large length scales, we may assume that the net result of
the confinement is that the membrane experiences an effective harmonic
potential which can be introduced as an additional term in the
Helfrich Hamiltonian:
\begin{equation}
{\cal H} = \int\,\frac{1}{2}\left[
\kappa\left(\nabla^2 h \right)^2
+\gamma h^2
\right]
d^2\vec{r},
\label{eq:helfrich2} 
\end{equation}
where $\gamma$ is a constant which will be determined later, and $h$
is measured from the mid-plane between the walls. In Fourier space
the energy reads
\begin{equation}
{\cal H} = \frac{l^4}{L^2}\sum_{\vec{q}}
\frac{1}{2}\left[\kappa q^4+\gamma\right]|h_q|^2,
\label{eq:qhelfrich2}  
\end{equation}
and the spectrum of fluctuations is given by
\begin{equation}
\langle|h_q|^2\rangle =\frac{k_BT L^2}{l^4\left(\kappa q^4+\gamma\right)}.
\label{eq:equipartition2}
\end{equation}
If $\gamma\gg \kappa(2\pi/L)^4$ (which is always satisfied for
sufficiently large $L$), then the mean-square fluctuation of the
height is given by
\begin{equation}
\Delta^2=\langle h\left(\vec{r}\right)^2\rangle=\left(\frac{l}{L}\right)^4
\sum_{\vec{q}} \langle |h_q|^2 \rangle =\frac{k_BT}{2\pi\sqrt{\kappa\gamma}}.
\label{eq:meanhfluct2}
\end{equation}
This result can be used for determining the value of $\gamma$. Each
point of the membrane has equal probability to be found anywhere
between the walls. Therefore,
\begin{equation}
\Delta^2=\frac{d^2}{12},
\label{eq:meanhfluct3}
\end{equation}
and by comparing Eqs.~(\ref{eq:meanhfluct2}) and
(\ref{eq:meanhfluct3}), we find that
\begin{equation}
\gamma=\frac{36(k_BT)^2}{\pi^2\kappa d^4}.
\label{eq:gamma1}
\end{equation}

Due to the thermal fluctuations, the membrane collides with the
confining walls and lose configurational entropy in these
collisions. The walls will therefore experience a disjoining
potential. Tracing over $|h_q|$ in Eq.(\ref{eq:qhelfrich2}) leads to
the following expression for the Helmholtz free energy
\begin{equation}
F=\frac{k_BT}{2}\sum_q \ln\left[\frac{\lambda^2l^4\left
(\kappa q^4+\gamma\right)}{2\pi L^2 k_BT}\right],
\label{eq:freeenergy}
\end{equation}
where $\lambda$ is the thermal de-Broglie wavelength of a microscopic
membrane patch of area $l^2$. The disjoining pressure between the
walls is then calculated by using
Eqs.~(\ref{eq:equipartition2})--(\ref{eq:freeenergy}) as follows:
\begin{equation}
p=-\frac{1}{L^2}\frac{\partial F}{\partial d}
=-\frac{1}{L^2}\frac{\partial F}{\partial \gamma}\frac{\partial \gamma}
{\partial d}=\frac{\partial \gamma}{\partial d}\frac{\Delta^2}{2}=
\frac{6(k_BT)^2}{\pi^2\kappa d^3}.
\label{eq:presssure}
\end{equation}
From this result it follows that the effective disjoining potential
per unit area \cite{safran_book}:
\begin{equation}
V=-\int_{\infty}^{d}\,p(x)dx=\frac{3(k_BT)^2}{\pi^2\kappa d^2},
\label{eq:potential}
\end{equation}
decays quadratically with $d$. 

\section{Membrane fluctuations near a single plane surface}
\label{sec:fluctuations}

\begin{figure}
\vspace{1.5cm}
  {\centering \hspace{1.5cm}\epsfig{file=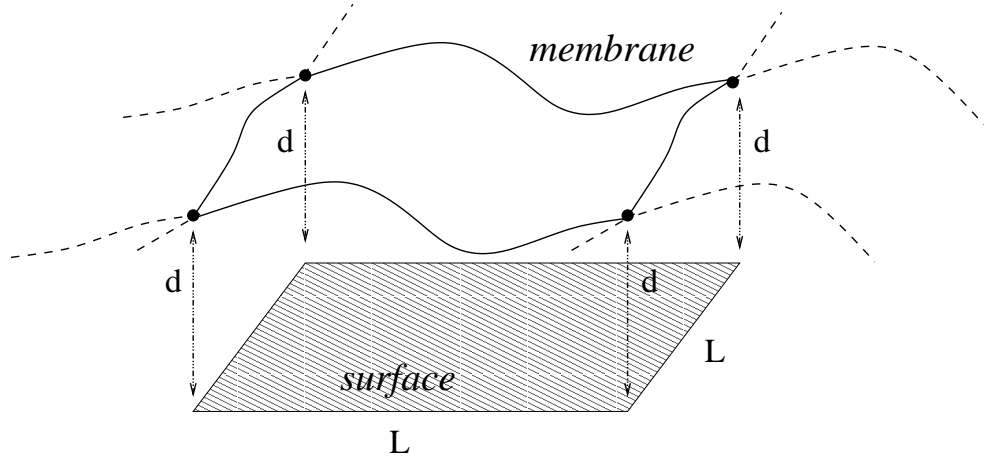,width=13cm}}
\caption{A square membrane fluctuating above a flat, impenetrable, surface. 
The function $h(x,y)$ denotes the height of the membrane above the
underlying surface. At the four corners of the surface
$h(0,0)=h(0,L)=h(L,0)=h(L,L)=d$. Outside the frame region, $h(x,y)$ is
defined by periodic extension.}
\label{fig:onewall_schm}
\end{figure}

\begin{figure}
\vspace{1.5cm}
  {\centering \hspace{1.5cm}\epsfig{file=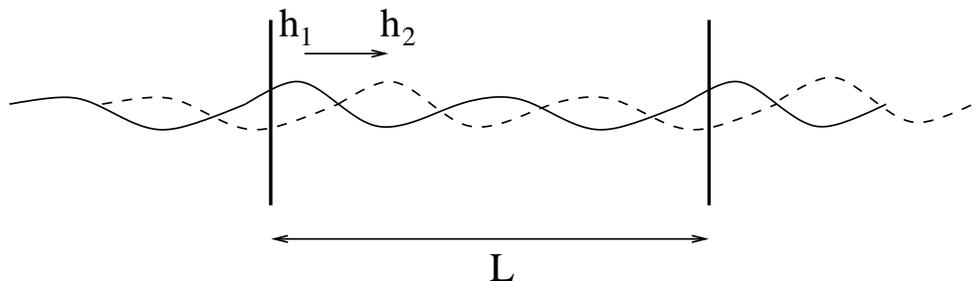,width=13cm}}
\caption{Configurations that are represented by the height functions 
$h_1(x,y)$ (solid curve) and $h_2(x,y)$ (dashed curve) belong to the
same sub-space of configurations if these configurations are invariant
under translation in the $x$-$y$ plane, i.e., $h_1(x,y)=h_2(x+a,y+b)$
with $0<a,b<L$.}
\label{fig:translation}
\end{figure}

Eq.(\ref{eq:potential}) for the disjoining potential has been derived
for the case of a membrane fluctuating between two walls. Does this
result also hold in the case of a membrane fluctuating near a {\em
single}\/ rigid wall? Consider a square membrane of linear size $L$
with bending rigidity $\kappa$ whose four corners are held a distance
$d$ above a flat, impenetrable, surface
(Fig.~\ref{fig:onewall_schm}). The height of the membrane relative to
the underlying surface is denoted by the function $h(x,y)$. In what
follows, we shall assume that $h(x,y)$ is periodic (with period $L$)
along both $x$ and $y$ directions. The pressure due to the collisions
between the fluctuating membrane and the surface must be
repulsive. However, it is not a-priori obvious why $p$ should be
proportional to $d^{-3}$, as predicted by
Eq.(\ref{eq:presssure}). Moreover, it is not even intuitively clear
whether this pressure should enhance or suppress the amplitude of the
membrane thermal fluctuations. The pinning of the edges of the
membrane and the EV interactions with the surface represent a
combination of attractive and repulsive potentials whose net effect is
not really well understood. A better understanding of this issue can
be obtained by comparing the configurational phase space of our model
system membrane with that of a freely fluctuating membrane. In the
free membrane case, we consider the ensemble of configurations for
which the {\em spatial}\/ average of the height,
$\bar{h}\equiv(1/L^2)\int h(x,y)\,dxdy$, is equal to some fixed
arbitrary value $c$. Setting $\bar{h}=c$ is necessary to avoid
multiple counting of physically equivalent configurations invariant
under a vertical translation along the $z$ direction. (Note the
difference in notation between the spatial average
$\overline{A(h)}\equiv (1/L^2)\int dxdy\,A(h)$ which is calculated for
a specific configuration, and $\langle A(h)\rangle$ which is the
statistical mechanical average over the ensemble of all possible
(distinct) configurations.) The phase space of free membrane
configurations can be further divided into sub-spaces, where the
height functions $h_1$ and $h_2$ of each two configurations included
in the same sub-space can be related by $h_1(x,y)=h_2(x+a,y+b)$ with
$0<a,b<L$. In other words, all the configurations in each sub-space
can be transformed into each other by a horizontal translation in the
$x-y$ plane (see Fig.~\ref{fig:translation}). This transformation does
not change $\bar{h}$ and, therefore, does not exclude (introduce)
allowed (forbidden) configurations from (into) the phase space of
configurations with $\bar{h}=c$. Since the Helfrich energy
[Eq.(\ref{eq:helfrich})] is invariant under translations, all the
configurations within each sub-space have exactly the same statistical
weight. The partition function which involves summing over all
possible configurations can be presented as summation over the
``sub-spaces'':
\begin{equation}
Z=\sum_{\rm configurations}e^{-\frac{{\cal H}}{k_BT}}=
\sum_{\rm sub-spaces}\Omega\, e^{-\frac{{\cal H}}{k_BT}}=
\sum_{\rm sub-spaces} e^{-\frac{(-k_BT\ln \Omega+{\cal H})}{k_BT}},
\label{eq:parti_new1}
\end{equation}
where $\Omega$ is the number of configurations included in each
sub-space.  The last equality in the above equation can be understood
as if each sub-space of the configurational phase space is represented
by only one configuration with height function $h(x,y)$ whose energy
is given by by the sum of Helfrich elastic energy and an extra term
that accounts for the ``degeneracy'' of the corresponding sub-space:
\begin{equation}
{\cal H}_{\rm sub-space}= -k_BT\ln\Omega+\int\,
\frac{\kappa}{2}\left(\nabla^2 h \right)^2d^2\vec{r}.
\label{eq:helfrich3} 
\end{equation} 
In the case of a free membrane, the number of configurations in each
sub-space is obviously the same: $\Omega=\Omega_0=(L/l)^2$. (Note that
for the purpose of counting the number of configurations, we
henceforth assume that two configuration are distinct only if they are
shifted by at least one grid cell of microscopic area $l^2$ with
respect to each other.)

\begin{figure}
\vspace{1.5cm}
  {\centering \hspace{1.5cm}\epsfig{file=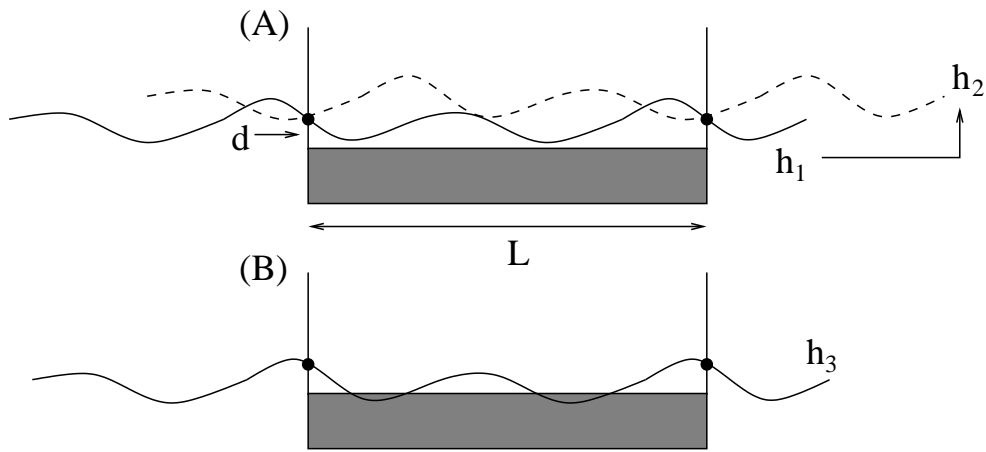,width=13cm}}
\caption{(A) The transformation between the configurations represented by the 
height functions $h_1(x,y)$ and $h_2(x,y)$ involves (i) a horizontal
translation in the $x-y$ plane and (ii) a vertical translation in the
normal $z$ direction that sets the corner at
$h_1(0,0)=h_2(0,0)=d$. (B)~Configurations such as the one represented
by the function $h_3$ intersect the underlying surface and, therefore,
should be excluded from the phase space.}
\label{fig:translation2}
\end{figure}

Let us repeat the above argument for our model system shown in
Fig.~\ref{fig:onewall_schm}. In this case, the mapping transformation
between configurations belonging to the same sub-space involves two
steps: (i) a horizontal translation in the $x-y$ plane, and (ii) a
vertical translation in the normal $z$ direction which sets the height
of the corners to be $h(0,0)=h(0,L)=h(L,0)=h(L,L)=d$ above the
underlying surface (see Fig.~\ref{fig:translation2}~(A)). The Helfrich
energy is invariant under these transformations. However, the vertical
translation may lead to the intersection of the membrane with the
surface and, therefore, to the exclusion of the configuration from the
sub-space of allowed configurations
(Fig.~\ref{fig:translation2}~(B)). The number of configurations left
in each sub-space $\Omega=\Omega(h(x,y),d)\leq\Omega_0$. Introducing
the function $G(h(x,y),d)\leq 1$, we can write
$\Omega=\Omega_0G(h(x,y),d)$ and rewrite Eq.(\ref{eq:helfrich3})
\begin{equation}
{\cal H}_{\rm sub-space}= -k_BT\ln\Omega_0
-k_BT\ln G\left(h\left(x,y\right),d\right)+\int\,
\frac{\kappa}{2}\left(\nabla^2 h \right)^2d^2\vec{r}.
\label{eq:helfrich4} 
\end{equation}
Adding the term $+k_BT\ln\Omega_0$ to Eq.(\ref{eq:helfrich4}) allows
us to replace the summation over ``sub-spaces'' back with summation
over all the possible configurations of a {\em ``free''} membrane
(without a surface). The effective Hamiltonian of this ``free''
membrane is given by
\begin{equation}
{\cal H}= -k_BT\ln G(h(x,y),d)+\int\,
\frac{\kappa}{2}\left(\nabla^2 h \right)^2d^2\vec{r}.
\label{eq:helfrich5} 
\end{equation}
There is no EV term in this Hamiltonian (since the membrane is assumed
to be free), but these interactions between the membrane and the
surface are properly accounted for by the first term on the right hand
side which quantifies the effect of the surface on the configurational
entropy of the membrane.

An interesting and unexpected result can be obtained in the $d=0$
limit, i.e., when the corners of the membrane are placed directly on
the surface. In this limit, the transformation defined between
configurations within each sub-space (see
Fig.~\ref{fig:translation2}~(A)) will almost always generate
``forbidden'' configurations that intersect the surface
(Fig.~\ref{fig:translation2}~(B)). Only when the pinning point
coincides with the global minimum of $h(x,y)$, then the membrane will
be positioned above the surface over the entire frame
region. Therefore, each sub-space includes only one configuration:
$\Omega=1$ (neglecting the measure-zero set of configurations with
multiple global minima), which means that
$G(h(x,y),0)=\Omega/\Omega_0=(l/L)^2$ does not dependent on
$h(x,y)$. Substituting this result into Eq.(\ref{eq:helfrich5}), we
find that a constant term was added to the Helfrich energy of the free
membrane. Therefore, the statistical mechanical properties of the
pinned membrane are identical to those of the free membrane and, in
particular, its fluctuation spectrum is also given by
Eq.(\ref{eq:equipartition}).

\begin{figure}
\vspace{1.5cm}
  {\centering \hspace{1.5cm}\epsfig{file=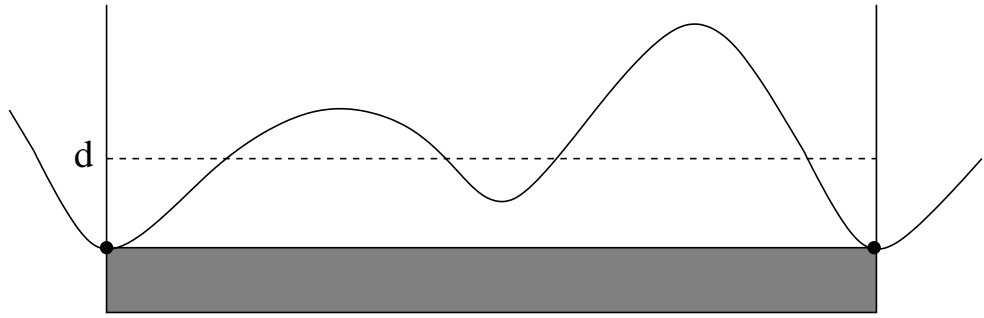,width=13cm}}
\caption{The membrane can be pinned to the surface without intersecting it 
only at points for which $h(x,y)-h_{\rm min}<d$, i.e., the membrane
points located below the horizontal dashed line in the
figure. Specifically, for $d=0$ the only possible pinning point is at
the global minimum of the function $h$.}
\label{fig:translation3}
\end{figure}

How can we evaluate the function $G(h(x,y),d)$ defined above? Let us
consider again the mapping transformation between configurations
belonging to the same configurational sub-space
(Fig.~\ref{fig:translation2}~(A)). This transformation changes the
pinning point by translating the membrane both horizontally and
vertically. The number of allowed configurations in the sub-space is
determined by the number of points on the membrane which can be placed
a height $d$ above the surface without causing any part of the
membrane to intersect the surface. The set of such possible pinning
points includes all the points on the membrane for which
$h(x,y)-h_{\rm min}\leq d$, where $h_{\rm min}$ is the global minimum
of the height function. These points are located below the dashed
horizontal line in Fig.~\ref{fig:translation3}. Denoting by
$A_p(h(x,y),d)<L^2$ the total projected area associated with this set
of possible pinning points, the function $G(h(x,y),d)=A_p/L^2$
represents the fraction of membrane points that satisfy the ``pinning
condition'' $h(x,y)-h_{\rm min}\leq d$.

Let us introduce the height distribution function of the membrane,
$p_{h(x,y)}(z)$. For a given height function $h(x,y)$,
$p_{h(x,y)}(z)dz$ gives the fraction of membrane points for which
$z<h(x,y)<z+dz$. The function $G(h(x,y),d)$ is the cumulative
distribution function associated with $p_{h(x,y)}(z)$:
\begin{equation}
G(h(x,y),d)=\int_{-\infty}^{h_{\rm min}+d}p_{h(x,y)}(z)dz.
\label{eq:distribution}
\end{equation} 
We proceed by approximating $p_{h(x,y)}(z)$ by a Gaussian distribution
function \cite{remark1}
\begin{equation}
p_{h(x,y)}(z)\sim 
\frac{1}{\sqrt{2\pi}\Delta_{h(x,y)}}\exp\left(\frac{(z-\bar{h})^2}
{2\Delta_{h(x,y)}^2}\right),
\label{eq:gaussdistribution}
\end{equation} 
where 
\begin{equation}
\Delta_{h(x,y)}^2\equiv\overline{(h-\overline{h})^2}
=\left(\frac{l}{L}\right)^4
\sum_{\vec{q}\neq 0} |h_q|^2.
\label{eq:meanhfluct4}
\end{equation}
Using Eq.(\ref{eq:gaussdistribution}) in Eq.(\ref{eq:distribution}),
we find :
\begin{equation}
G(h(x,y),d)\sim \frac{1}{2}\left[1+{\rm erf}\left(
\frac{h_{\rm min}-\bar{h}+d}{\sqrt{2}\Delta_{h(x,y)}}\right)\right]=
\frac{1}{2}\left[1+{\rm erf}\left(-\alpha+
\frac{d}{\sqrt{2}\Delta_{h(x,y)}}\right)\right],
\label{eq:distribution2}
\end{equation}
where ${\rm erf}(x)=(2/\sqrt{\pi})\int_0^x e^{-u^2}du$ is the
standard error function, and $\alpha\equiv(\bar{h}-h_{\rm
min})/\sqrt{2}\Delta_{h(x,y)}>0$. The function $G(h(x,y),d)$ given by
Eq.(\ref{eq:distribution2}) satisfies the boundary condition that
$G(d\rightarrow\infty)\rightarrow 1$. The value of $\alpha$ can be set
by imposing the other boundary condition (see discussion above) that 
$G(d=0)=(l/L)^2$, which gives
\begin{equation}
\alpha={\rm erf}^{-1}\left[1-\left(\frac{2l^2}{L^2}\right)\right].
\label{eq:alpha}
\end{equation}

Using the Fourier representation $\{h_q\}$ of the function $h(x,y)$
and Eqs.~(\ref{eq:helfrich5}) and (\ref{eq:distribution2}), we find
that the statistical mechanical properties of the pinned membrane can
be derived by considering a free membrane whose thermal behavior is
governed by the Hamiltonian:
\begin{equation}
{\cal H}= -k_BT\ln\left[
\frac{1}{2}+\frac{1}{2}{\rm erf}\left(-\alpha+
\frac{d}{\sqrt{2}\Delta_{h(x,y)}}\right)\right]+
\frac{l^4}{L^2}\sum_{\vec{q}}\frac{\kappa}{2} q^4|h_q|^2,
\label{eq:helfrich6}
\end{equation}
where $\alpha$ is given by Eq.(\ref{eq:alpha}). The dependence of the
first term on the right hand side of Eq.(\ref{eq:helfrich6}) on
$\{h_q\}$ is contained in the variable $\Delta_{h(x,y)}$ (see
Eq.(\ref{eq:meanhfluct4})). The spectral intensity of the pinned
membrane can be calculated ed by using the equipartition theorem:
\begin{equation}
k_BT=\left \langle h_q\frac{\partial {\cal H}}{\partial h_q}
\right\rangle
\label{eq:equipartition3}
\end{equation}
Introducing the variable $d^*=-\alpha+d/\sqrt{2}\Delta_{h(x,y)}$, the
right hand side of Eq.(\ref{eq:equipartition3}) can be written as
follows:
\begin{eqnarray}
h_q\frac{\partial {\cal H}}{\partial h_q}&=&\left(\frac{l^4}{L^2}\right)
\kappa q^4 |h_q|^2-k_BT\left(\frac{2}{\sqrt{\pi}}\frac{e^{-(d^*)^2}}
{\left[1+{\rm erf}\left(d^*\right)\right]}\right)
\frac{\partial d^*}{\partial h_q}h_q
\nonumber \\
&=&\left(\frac{l^4}{L^2}\right)
\kappa q^4 |h_q|^2+k_BT
\left(\frac{2}{\sqrt{\pi}}\frac{e^{-(d^*)^2}}
{\left[1+{\rm erf}\left(d^*\right)\right]}\right)
\left(\frac{d}
{\sqrt{2}\Delta_{h(x,y)}^2}\right) \frac{\partial \Delta_{h(x,y)}}
{\partial h_q}h_q
\nonumber \\
&=&\left(\frac{l^4}{L^2}\right)
\left\{
\kappa q^4 +\left(k_BT\frac{d}{\Delta_{h(x,y)}^3L^2}\right)
\left(\sqrt{\frac{2}{\pi}}\frac{e^{-(d^*)^2}}
{\left[1+{\rm erf}\left(d^*\right)\right]}\right)
\right\}|h_q|^2.
\label{eq:equipartition4}
\end{eqnarray}
Eq.(\ref{eq:equipartition4}) represents a set of linear equations (one
for each Fourier mode $q\neq 0$). These equations are coupled to each
other through the mean-square height fluctuation $\Delta_{h(x,y)}$
(see Eq.(\ref{eq:meanhfluct4})) appearing both explicitly on the
second term in the curly brackets, as well as in the definition of the
variable $d^*$. For both $d\rightarrow \infty$ (free membrane) and
$d=0$ (membrane pinned directly to the surface) the second term in the
curly brackets vanishes and, therefore, Eq.(\ref{eq:equipartition})
which describes the fluctuation spectrum of a free membrane is
recovered in these two limits, as argued above. For finite values of
$d$, a further approximation can be made by replacing the spatial
average $\Delta_{h(x,y)}$ with $\Delta_0$, the ensemble average over
free membrane configurations (see Eq.(\ref{eq:meanhfluct})). This
approximation leads to the decoupling of the set of equations
(\ref{eq:equipartition4}) and yields the following result:
\begin{equation}
\langle|h_q|^2\rangle =\frac{k_BT L^2}{l^4\left[
\kappa q^4 +\left(k_BT\frac{d}{\Delta_0^3L^2}\right)
\left(\sqrt{\frac{2}{\pi}}\frac{e^{-(d^*)^2}}
{\left[1+{\rm erf}\left(d^*\right)\right]}\right)
\right]},
\label{eq:equipartition5}
\end{equation}
where within the approximation of replacing $\Delta_{h(x,y)}$ with
$\Delta_0$, we also set
$d^*=-\alpha+d/\sqrt{2}\Delta_0$. Eq.(\ref{eq:equipartition5}) has the
same form as Eq.(\ref{eq:equipartition2}) which describes the
power spectrum of a membrane fluctuating under the action of a
uniform harmonic potential of strength
\begin{equation}
\gamma_{\rm eff}=
\left(k_BT\frac{d}{\Delta_0^3L^2}\right)
\left(\sqrt{\frac{2}{\pi}}
\frac{e^{-\left(-\alpha+\frac{d}{\sqrt{2}\Delta_0}\right)^2
}}{\left[1+{\rm
erf}\left(-\alpha+\frac{d}{\sqrt{2}\Delta_0}\right)\right]}\right).
\label{eq:gamaeff}
\end{equation}
From Eq.(\ref{eq:equipartition5}) we conclude that the thermal height
fluctuations of the pinned membraned are attenuated compared to the
fluctuations of a free membrane. The mean square fluctuation amplitude
of a mode with wavevector $q=2\pi n/L$ is reduced by a factor
\begin{equation}
I_n\equiv\frac{\langle|h_q(\gamma_{\rm eff})|^2\rangle}
{\langle|h_q(\gamma_{\rm eff}=0)|^2\rangle}=\frac{\kappa q^4}{\kappa
q^4 +\gamma_{\rm eff}}=\frac{n^4}{n^4+n_{\gamma}^4},
\label{eq:factor}
\end{equation}
where 
\begin{equation}
n_{\gamma}^4=\left(\frac{\gamma_{\rm eff}}{\kappa}\right)
\left(\frac{L}{2\pi}\right)^4=\frac{k_BT}{(2\pi)^4\kappa}
\left(\frac{dL^2}{\Delta_0^3}\right)
\left(\sqrt{\frac{2}{\pi}}
\frac{e^{-\left(-\alpha+\frac{d}{\sqrt{2}\Delta_0}\right)^2
}}{\left[1+{\rm
erf}\left(-\alpha+\frac{d}{\sqrt{2}\Delta_0}\right)\right]}\right).
\label{eq:ngamma}
\end{equation}
is a dimensionless number that governs the crossover between the
regimes of {\em damped}\/ ($n^4\ll n_{\gamma}^4)$ and {\em free}\/
($n^4\gg n_{\gamma}^4)$ thermal fluctuations. Values of $n_{\gamma}^4$
are plotted in Fig.~\ref{fig:ngamma} for different values of $d$ and
for $\kappa=10k_BT$ and $L=10l\sim 50\ {\rm nm}$ ($\alpha=1.645$ - see
Eq.(\ref{eq:alpha})). The maximum value of $n_{\gamma}^4\sim 0.34$ is
obtained for $d\sim 0.025-0.04 L\sim 1.25-2\ {\rm nm}$. This value of
$n_{\gamma}^4\sim 0.34$ is too small to have any noticeable effect on
the spectrum of thermal fluctuations, except for the largest mode
($n=1$) whose square amplitude is suppressed by a factor of $I_1\sim
1/1.34=0.75$. In comparison, the square amplitudes of second and third
largest modes are reduced by only factors of $I_{\sqrt{2}}\sim 0.92$
and $I_{2}\sim 0.98$, respectively. We also observe from
Fig.~\ref{fig:ngamma} that for smaller values of $\kappa$, the maximum
values of $n_{\gamma}$ occurs at larger values of $d/L$. This behavior
is anticipated since the smaller $\kappa$ the larger the amplitude of
the thermal fluctuations and, therefore, the greater the range of
steric repulsion between the membrane and the surface. However, even
for very soft membranes with $\kappa=3k_BT$, the maximum is still
reached at $d/L<0.1$, i.e., only a few nanometers above the surface.

\section{Computer simulations}

\begin{figure}
\vspace{1.5cm}
  {\centering \hspace{1.5cm}\epsfig{file=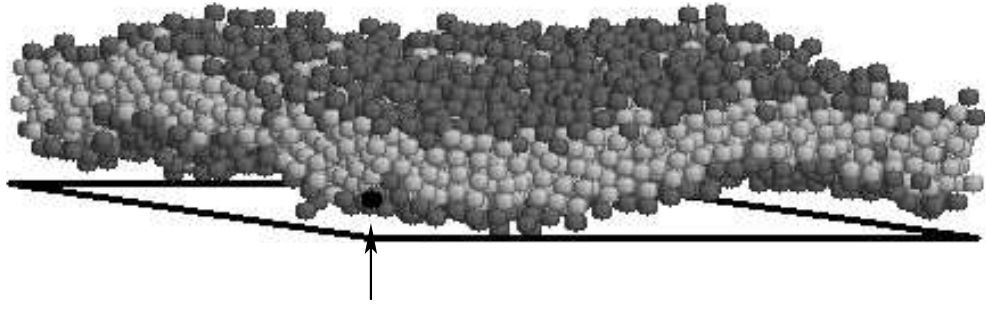,width=13cm}}
\caption{Equilibrium configuration of a membrane consisting of 2000 
lipids. Each lipid is represented by a trimer of one ``hydrophilic''
bead (dark gray sphere) and two ``hydrophobic'' beads (light gray
spheres). The membrane is fluctuating above a plane surface (frame
indicated by a thick black line), while the position of the center of
one of the hydrophilic beads (appearing at the front of the figure and
indicated by the black sphere and an arrow) is fixed at
$\vec{r}=(x,y,z)=(0,0,d+\sigma/2)$.}
\label{fig:configuration}
\end{figure}

One of the important recent advances in soft-matter simulations is the
development of coarse-grained (CG) bilayer membrane models in which
the membranes are simulated without direct representation of the
embedding solvent \cite{farago:2003,Brannigan_Brown}. These implicit
solvent ("solvent free") CG models require modest CPU and memory
resources and, therefore, can be used for simulations of
mesoscopically large membranes over long enough timescales to address
experimental reality. Here, we use an implicit solvent CG model to
test the validity and accuracy of the analytical predictions discussed
above. A snapshot from the simulations is shown in
Fig.~\ref{fig:configuration}. Each lipid molecule is represented by a
short string of three spherical beads of diameter $\sigma$, where one
of the beads (depicted as a dark gray sphere in
Fig.~\ref{fig:configuration}) represents the hydrophilic head group
and two beads (light gray spheres in Fig.~\ref{fig:configuration})
represent the hydrophobic tail of the lipid. The details of the model
and of the molecular simulations are given in refs.~\cite{cooke:2005}
and \cite{farago:2008}, including the description of a new Monte Carlo
scheme (Mode Excitation Monte Carlo) which has been applied to
accelerate the relaxation of the five largest Fourier modes
corresponding to $n=1,\sqrt{2},2,\sqrt{5},\sqrt{8}$. The simulated
system consists of a bilayer membrane of 2000 lipids and a surface
(whose boundary is indicated by a thick black line) located at $z=0$
below which the beads cannot be found. The {\em center}\/ of the head
bead of one of the lipids in the lower leaflet (indicated by a black
sphere appearing at the front of Fig~\ref{fig:configuration}) is held
fixed at $\vec{r}=(x,y,z)=(0,0,d+\sigma/2)$. Note that in our
simulations, $d$ is defined as the distance between the surface and
the {\em bottom}\/ of the particle whose position is held fixed. The
simulations were carried out on the ``high performance on demand
computing cluster'' at Ben Gurion University. For each value of $d$,
the simulation results appearing below are based on 16 independent
runs, each of $1.2\times 10^{6}$ Monte Carlo time units. The first
$2\times 10^{5}$ time units of each run were discarded from the
statistical analysis.

\begin{figure}
\vspace{1.5cm}
  {\centering \hspace{1.5cm}\epsfig{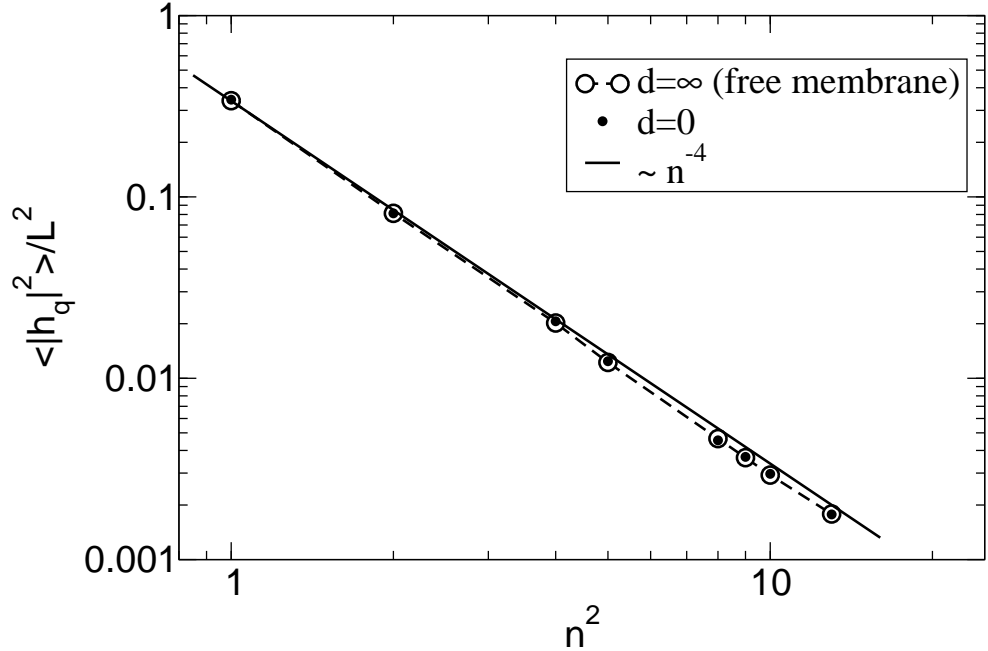}}
\caption{The fluctuation spectrum of a membrane of N = 2000 lipids. 
The results for $d=0$ (membrane pinned directly to the surface) are
shown by small solid circles. These results are essentially identical
to those obtained from simulations of a free membrane ($d=\infty$),
which are represented by larger open circles connected
with a dashed line. The solid line indicates the asymptotic
$\langle|h_q|^2\rangle\sim n^{-4}$ power law.}
\label{fig:spectrum}
\end{figure}

The simulations were conducted in the constant surface tension
ensemble, at vanishing surface tension. We set the bead diameter,
$\sigma\sim 5/6\ {\rm nm}$, so that the bilayer membrane thickness is
$l\sim 6\sigma\sim 5\ {\rm nm}$, and the (mean) linear size of the
bilayer $L=(36.4\pm 0.1)\,\sigma\sim 30\ {\rm nm}$ (for all values of
$d$). The spectrum of the membrane height fluctuations $\langle
|h_q|^2\rangle$ is plotted in Fig.~\ref{fig:spectrum} for both $d=0$
and $d=\infty$ (i.e., for a freely fluctuating membrane). The
computational results fully confirm our analytical prediction that the
power spectra in these two cases are identical. By fitting the
computational results to the asymptotic form (for small values of $n$)
Eq. (\ref{eq:equipartition}), we obtain that the bending rigidity of
the membrane $\kappa\sim (7.8\pm 0.2)\ k_BT$.

For $d\neq 0$ we expect the power spectrum of the membrane to be
``almost'' identical to the power spectrum of the free membrane. More
precisely, Eq.~(\ref{eq:factor}) predicts that the mean square
fluctuation amplitudes of the modes will be reduced by a factor $I_n$
which, except for the longest $n=1$ mode, is very close to unity. This
prediction is very well supported by our computational results which
are summarized in Table~\ref{tab:factors}. The table gives the values
$I_1$ and $I_{\sqrt{2}}$ (corresponding, respectively, to the largest
and second largest Fourier modes) for different values of $d$. Within
the statistical accuracy of our simulation results, we found no change
in the fluctuation amplitudes of all the other modes corresponding to
wavenumbers $n\geq 2$. In order to evaluate the quantitative agreement
between the computational data presented in Table~\ref{tab:factors}
and the above analytical theory, we use Eqs.~(\ref{eq:meanhfluct}),
(\ref{eq:alpha}), (\ref{eq:factor}), and (\ref{eq:ngamma}) to
calculate the factor $I_1$ for different values of $d$ and for the set
of parameters relevant to our simulations: $\kappa=7.8 k_BT$,
$L=6l=36\sigma$. The results of the calculation along with our
computational results (Table~\ref{tab:factors}) are plotted in
Fig.~\ref{fig:factor}. As can clearly be seen in the figure, the
agreement between the analytical and computational results is quite
good. This agreement lends support for the validity and accuracy of
our theoretical analysis of the fluctuation spectrum.

\section{The disjoining pressure}

\begin{figure}
\vspace{1.5cm}
  {\centering \hspace{1.5cm}\epsfig{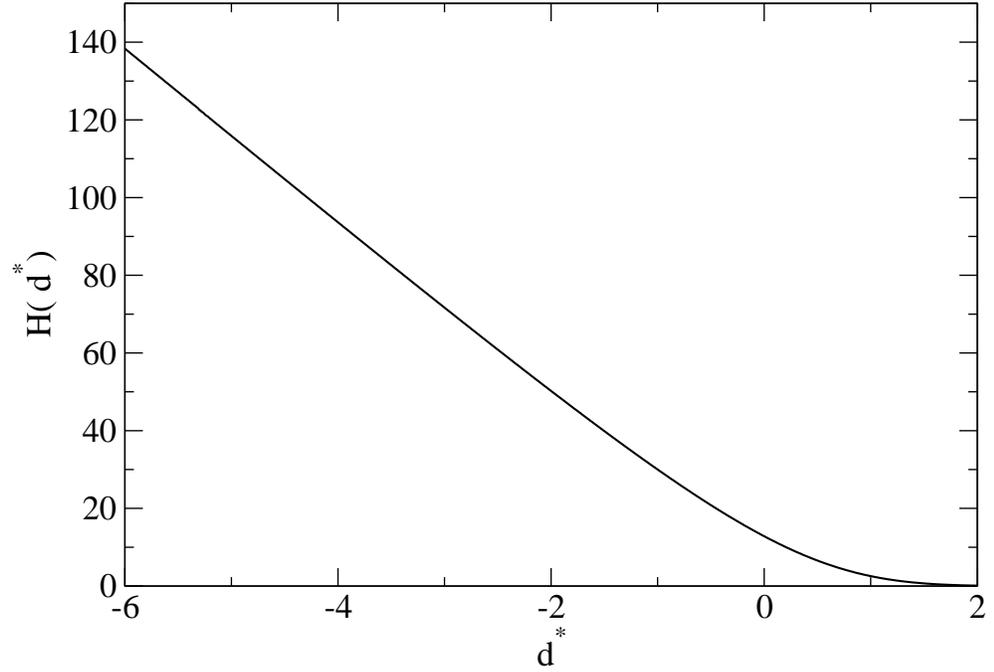}}
  \caption{The scaling function $H(d^*)$ (see definition -
  Eq.(\ref{eq:pressure4})).}  
\label{fig:scale} 
\end{figure}

The fact that surface influence on the bilayer fluctuations can be
effectively described in terms of a uniform harmonic potential, does
not imply that the disjoining pressure between the surface and the
membrane follows Eq.(\ref{eq:presssure}). The pressure can be related
to Hamiltonian (\ref{eq:helfrich6}) through the following equation:
\begin{equation}
p=-\frac{1}{L^2}\left\langle\frac{\partial{\cal H}}{\partial d}\right\rangle,
\label{eq:pressure2}
\end{equation}
from which we readily derive that
\begin{equation}
p=\frac{k_BT}{L^2}
\left\langle\frac{1}{\Delta_{h(x,y)}}
\left(\sqrt{\frac{2}{\pi}}\frac{e^{-\left(-\alpha+\frac{d}
{\sqrt{2}\Delta_{h(x,y)}}
\right)^2}}
{\left[1+{\rm
erf}\left(-\alpha+\frac{d}{\sqrt{2}\Delta_{h(x,y)}}\right)\right]}\right)
\right\rangle.
\label{eq:pressure3}
\end{equation} 
Following the approach described in section \ref{sec:fluctuations},
the thermal average in Eq.(\ref{eq:pressure3}) can be approximately
evaluated by replacing $\Delta_{h(x,y)}$ with $\Delta_0$
(Eq.(\ref{eq:meanhfluct})), which gives the following expression:
\begin{equation}
p=\frac{k_BT}{L^2\Delta_0}
\left(\sqrt{\frac{2}{\pi}}
\frac{e^{-\left(-\alpha+\frac{d}{\sqrt{2}\Delta_0}\right)^2
}}{\left[1+{\rm
erf}\left(-\alpha+\frac{d}{\sqrt{2}\Delta_0}\right)\right]}\right)=
p^*
\left(
\frac{Ce^{-\left(d^*\right)^2}}{\left[1+{\rm erf}\left(d^*\right)\right]}
\right)\equiv p^*H\!\left(d^*\right),
\label{eq:pressure4}
\end{equation}
where $p^*=(k_BT\kappa)^{1/2}/L^3$, $d^*=-\alpha+d/\sqrt{2}\Delta_0$,
and $C=(2\pi)^2(2/6.03\pi)^{1/2}=12.83$. The scaling function
$H(d^*)$, which is plotted in Fig.~\ref{fig:scale}, decreases
monotonically with $d^*$. At large distances ($d\rightarrow \infty$)
the pressure decreases as $\sim\exp[-(d/L)^2]$. The maximum
pressure occurs when the membrane is in direct contact with the
surface ($d=0$). The contact pressure, however, does not diverge but
rather reaches the following finite value $p(d=0)=p^*H(-\alpha)\simeq
p^*\sqrt{\pi}C\alpha$. Using Eqs.~(\ref{eq:alpha}) and
(\ref{eq:pressure4}), one can easily realize the contact pressure
{\em diminishes}\/ with the size of the membrane patch:
\begin{equation}
p(d=0)=p^*\sqrt{\pi}C\cdot{\rm erf}^{-1}\left[1-\left(2l^2/L^2\right)\right]
\sim \frac{\sqrt{k_BT\kappa}}{L^3}\ln\left(\frac{L}{l}\right).
\label{eq:pressure5}
\end{equation}

\section{Concluding remarks}

\begin{figure}
\vspace{1.5cm}
  {\centering \hspace{1.5cm}\epsfig{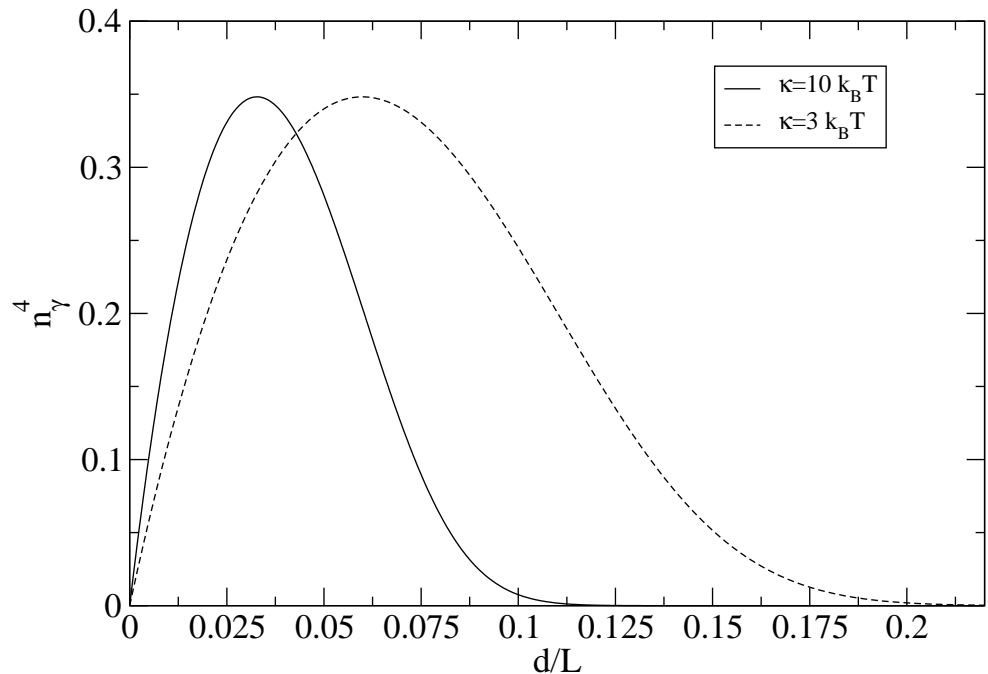}}
\caption{The dimensionless parameter $n_{\gamma}^4$ (see definition in 
text - Eq.{\protect \ref{eq:ngamma}}) as a function of the height of
the pinning points from the surface. The values have been calculated
for a membrane of linear size $L=10l$ with bending rigidity
$\kappa=10k_BT$ (solid line) and $\kappa=3K_BT$ (dashed line).}
\label{fig:ngamma}
\end{figure}

\begin{figure}
\vspace{1.5cm}
  {\centering \hspace{1.5cm}\epsfig{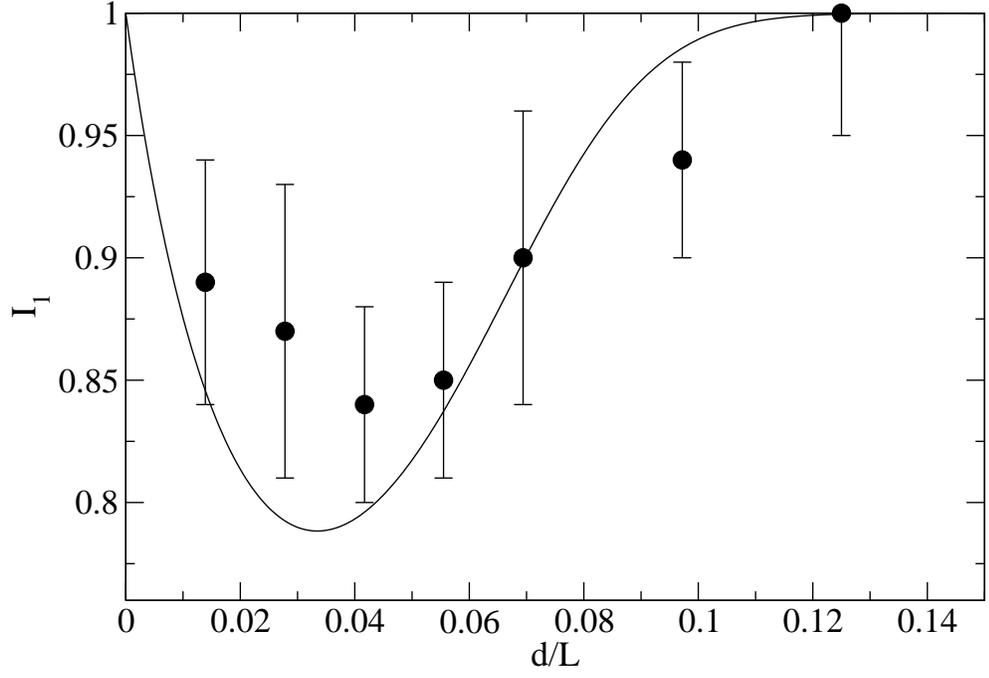}}
\caption{The factor $I_1$ as a function of $d$.  Solid circles -
computational results (see also Table~\ref{tab:factors}). Solid line -
analytical results for the computationally relevant parameters: $L=6l$
and $\kappa=7.8k_BT$.}
\label{fig:factor}
\end{figure}

In this paper, we analyzed the influence of EV volume effects on the
statistical mechanical properties of supported membranes. Using
analytical calculations and Monte Carlo simulations, we determined the
fluctuation spectrum of a bilayer membrane patch of a few tens of
nanometer in size whose corners are located at a fixed distance $d$
above a plane rigid surface. We found that the surface has influence
on the fluctuation spectrum of the membrane only when the pinning
distance $d$ is sufficiently small ($d\lesssim 0.1L$ - see
Figs.~\ref{fig:ngamma} and \ref{fig:factor}). This result can be
easily understood given the fact that the amplitude of the height
fluctuations of a free membrane patch satisfies $\Delta_0\sim 0.02L$
(see Eq.~(\ref{eq:meanhfluct})) and, obviously, the membrane hardly
collides with the underlying surface when $d\gg \Delta_0$. At small
distances ($d\simeq \Delta_0$), the surface influence on the bilayer
fluctuation spectrum resembles that of a uniform harmonic confining
potential of the form: $V=(1/2)\gamma_{\rm eff}(d) h^2$. Both
analytically and computationally we find that the strength
$\gamma_{\rm eff}(d)$ of the effective harmonic potential is extremely
small and has noticeable impact only on the amplitudes of the very
largest fluctuation modes. More remarkably, the confinement effect
vanishes ($\gamma_{\rm eff}(d)=0$), when the membrane is brought into
direct contact with the surface ($d=0$). This unexpected and
counterintuitive result can be explained by the fact that the thermal
motion of the membrane is not really confined within a finite spatial
region (as in a stack of bilayer membranes) but only restricted on one
side. Therefore, the primary effect of the collisions with the
underlying surface is to push the membrane ``upward'' rather than to
suppress the amplitude of the fluctuations. This is also the reason
why the disjoining pressure does not diverge when $d$ vanishes.

The significance of our findings should be considered in light of
previous theoretical attempts to quantify the steric effects between
membranes and underlying supported interfaces. Some of these studies
describe the wall-membrane pressure by means of
Eq.(\ref{eq:presssure}), which we have shown to be irrelevant for this
problem. One particular problem that should be reanalyzed in light of
our new results is the theoretical interpretation of the fluctuation
spectra of red blood cells~\cite{Zilker:1987}. The plasma membrane of
red blood cells is attached to a triangulated network of flexible
spectrin proteins with mesh size $\xi\sim 60-100\ {\rm nm}$. The
spectrum of red blood cells fluctuations was analyzed in terms of the
Helfrich Hamiltonian with both curvature and {\em scale dependent}\/
surface tension terms, where the latter term originates from the
coupling to the cytoskeleton. Using a Gaussian network model, Fournier
{\em et al.}\/~\cite{Fournier:2004,Fournier:2006} showed that the
effective surface tension exhibits a steep crossover from a
vanishingly small value at length scales smaller than $\xi$ to some
finite value at scales larger than $\xi$. Gov {\em et
al.}~\cite{Gov:2003,Gov_Safran:2004} argued that, in addition, a
uniform harmonic potential term must be introduced in the Helfrich
effective surface Hamiltonian, which accounts for the confinement
effect due to the steric repulsion between the spectrin and the
bilayer. Our statistical mechanical analysis partially supports this
phenomenological argument. On the one hand, our
Eq.(\ref{eq:equipartition5}) can be interpreted in terms of a uniform
harmonic potential that acts on the membrane. On the other hand, our
estimation of the strength of the effective harmonic potential makes
it questionable whether the origin of it can be attributed to EV
interactions (between the bilayer and the spectrin) alone. It seems
more likely to relate this additional confinement term to the
junctional complexes (of short actin filaments, globular band 4.1, and
other proteins) which connect the membrane to the cytoskeleton and
restrict the membrane height fluctuations around the points of
attachment.  We thus speculate that, just like the surface tension,
the strength of the effective harmonic potential $\gamma_{\rm eff}$
must also be scale dependent. At length scales below the mesh size, we
expect the value of $\gamma_{\rm eff}$ to be governed by EV effects
and, therefore, to be extremely small. Above the mesh size, the
strength of the harmonic confinement will be determined by the
strength of the periodic pinning of the membrane to the cytoskeleton
which, presumably, result in a larger value of $\gamma_{\rm eff}$. It
should be stressed here that there is currently no proof (or even a
reasoned argument) that the long wavelength fluctuations of red cells
are indeed harmonically confined. One should also bear in mind that on
the scale of the mesh size of the network, the problem is quite
intricate and issues such as connectivity defects and the motion of
the protein anchors must be properly addressed.

Discussions with Nir Gov, Thorsten Auth, Sam Safran and Phil Pincus
are gratefully acknowledged.


\newpage
\begin{table}
\begin{center}
\begin{tabular}{| r || l | l | }
    \hline
    $d$ & $I_1$ & $I_{\sqrt{2}}$ \\ \hline
    $0.5\sigma$ & $0.89\ (5)$ & $0.99\ (4)$ \\ \hline
    $   \sigma$ & $0.87\ (6)$ & $0.97\ (4)$ \\ \hline
    $1.5\sigma$ & $0.84\ (4)$ & $0.94\ (4)$ \\ \hline
    $  2\sigma$ & $0.85\ (4)$ & $0.95\ (4)$ \\ \hline
    $2.5\sigma$ & $0.90\ (6)$ & $0.96\ (4)$ \\ \hline
    $3.5\sigma$ & $0.94\ (4)$ & $0.99\ (4)$ \\ \hline
    $4.5\sigma$ & $1.00\ (5)$ & $1.00\ (4)$ \\ \hline
\end{tabular}
\end{center}
\caption{The factors $I_1$ and $I_{\sqrt{2}}$ (see Eq.(\ref{eq:factor})) by 
which the mean square fluctuation amplitudes of the largest ($n=1$)
and second largest ($n=\sqrt{2}$) modes are reduced as compared to the
square fluctuation amplitudes of a free membrane. The height $d$
denotes the distance between the bottom of the fixed spherical bead
and the underlying surface.}
\label{tab:factors}
\end{table}

\end{document}